\documentclass{elsart}

\usepackage{graphicx,amssymb}
\usepackage{unites2e}
\begin{document}

\begin{frontmatter}



\title{The First VERITAS Telescope}


\author[Leeds]{Holder, J.}
\author[Utah]{Atkins, R.W.}
\author[Tanta]{Badran, H.M.}
\author[UMass]{Blaylock, G.}
\author[Leeds]{Bradbury, S.M.}
\author[WashU]{Buckley, J.H.}
\author[Argonne]{Byrum, K.L.}
\author[ISU]{Carter-Lewis, D.A.}
\author[UCLA]{Celik, O.}
\author[UCLA]{Chow, Y.C.K.}
\author[UCD]{Cogan, P.}
\author[Purdue]{Cui, W.}
\author[UCD]{Daniel, M.K.}
\author[Oxford]{de la Calle Perez, I.}
\author[UCD]{Dowdall, C.}
\author[WashU]{Dowkontt, P.}
\author[Grinnell]{Duke, C.}
\author[Penn]{Falcone, A.D.}
\author[UCLA]{Fegan, S.J.}
\author[Purdue]{Finley, J.P.}
\author[Columbia]{Fortin, P.}
\author[Adler]{Fortson, L.F.}
\author[Whipple]{Gibbs, K.}
\author[NUI]{Gillanders, G.}
\author[Leeds]{Glidewell, O.J.}
\author[Leeds]{Grube, J.}
\author[WashU]{Gutierrez, K.J.}
\author[Adler]{Gyuk, G.}
\author[Utah]{Hall, J.}
\author[McGill]{Hanna, D.}
\author[Argonne,Chicago]{Hays, E.}
\author[Whipple]{Horan, D.}
\author[WashU]{Hughes, S.B.}
\author[Chicago]{Humensky, T.B.}
\author[ISU]{Imran, A.}
\author[WashU]{Jung, I.}
\author[Iowa]{Kaaret, P.}
\author[NUI]{Kenny, G.E.}
\author[Utah]{Kieda, D.}
\author[McGill]{Kildea, J.}
\author[Leeds]{Knapp, J.}
\author[WashU]{Krawczynski, H.}
\author[ISU]{Krennrich, F.}
\author[NUI]{Lang, M.J.}
\author[Utah]{LeBohec, S.}
\author[Chicago]{Linton, E.}
\author[Whipple]{Little, E.K.}
\author[Leeds]{Maier, G.}
\author[Utah]{Manseri, H.}
\author[Leeds]{Milovanovic, A.}
\author[GMIT]{Moriarty, P.}
\author[Columbia]{Mukherjee, R.}
\author[Leeds]{Ogden, P.A.}
\author[UCLA]{Ong, R.A.}
\author[WashU]{Perkins, J.S.}
\author[Purdue]{Pizlo, F.}
\author[ISU]{Pohl, M.}
\author[UCD]{Quinn, J.}
\author[McGill]{Ragan, K.}
\author[Cork]{Reynolds, P.T.}
\author[Whipple]{Roache, E.T.}
\author[Leeds]{Rose, H.J.}
\author[ISU]{Schroedter, M.}
\author[Purdue]{Sembroski, G.H.}
\author[ISU]{Sleege, G.}
\author[Adler]{Steele, D.}
\author[Chicago]{Swordy, S.P.}
\author[Leeds]{Syson, A.}
\author[NUI]{Toner, J.A.}
\author[McGill]{Valcarcel, L.}
\author[UCLA]{Vassiliev, V.V.}
\author[Chicago]{Wakely, S.P.}
\author[Whipple]{Weekes, T.C.}
\author[Leeds]{White, R.J.}
\author[UCSC]{Williams, D.A.}
\author[Argonne]{Wagner, R.}

\address[Leeds]{School of Physics and Astronomy, University of Leeds, Leeds, LS2 9JT, UK}
\address[Utah]{Physics Department, University of Utah, Salt Lake City, UT 84112, USA}
\address[Tanta]{Physics Department, Faculty of Science, Tanta University, Tanta 31527, Egypt}
\address[UMass]{Department of Physics, University of Massachussetts, Amherst, MA 01003-4525, USA}
\address[WashU]{Department of Physics, Washington University, St. Louis, MO 63130, USA}
\address[Argonne]{Argonne National Laboratory, 9700 S. Cass Avenue, Argonne, IL 60439, USA}
\address[ISU]{Department of Physics and Astronomy, Iowa State University, Ames, IA 50011, USA}
\address[UCLA]{Department of Physics and Astronomy, University of California, Los Angeles, CA 90095, USA}
\address[UCD]{School of Physics, University College Dublin, Belfield, Dublin 4, Ireland}
\address[Purdue]{Department of Physics, Purdue University, West Lafayette, IN 47907, USA}
\address[Oxford]{Department of Physics, University of Oxford, Oxford, OX1 3RH, UK}
\address[Grinnell]{Department of Physics, Grinnell College, Grinnell, IA 50112-1690, USA}
\address[Penn]{Department of Astronomy and Astrophysics, Penn State University, University Park, PA 16802, USA}
\address[Columbia]{Department of Physics and Astronomy, Barnard College, Columbia University, NY 10027}
\address[Adler]{Astronomy Department, Adler Planetarium and Astronomy Museum, Chicago, IL 60605, USA}
\address[Whipple]{Fred Lawrence Whipple Observatory, Harvard-Smithsonian Center for Astrophysics, Amado, AZ 85645, USA}
\address[NUI]{Physics Department, National University of Ireland, Galway, Ireland}
\address[McGill]{Physics Department, McGill University, Montreal, QC H3A 2T8, Canada}
\address[Chicago]{Enrico Fermi Institute, University of Chicago, Chicago, IL 60637, USA}
\address[Iowa]{Department of Physics and Astronomy, Van Allen Hall, Iowa City, IA 52242, USA}
\address[GMIT]{Department of Physical and Life Sciences, Galway-Mayo Institute of Technology, Dublin Road, Galway, Ireland}
\address[Cork]{Department of Applied Physics and Instumentation, Cork Institute of Technology, Bishopstown, Cork, Ireland}
\address[UCSC]{Santa Cruz Institute for Particle Physics and Department of Physics, University of California, Santa Cruz, CA 95064, USA}

\begin{abstract}

 The first atmospheric Cherenkov telescope of VERITAS (the Very Energetic
 Radiation Imaging Telescope Array System) has been in operation since February
 2005. We present here a technical description of the instrument and a summary
 of its performance. The calibration methods are described, along with the
 results of Monte Carlo simulations of the telescope and comparisons between real
 and simulated data. The analysis of TeV $\gamma$-ray observations of the Crab
 Nebula, including the reconstructed energy spectrum, is shown to give results
 consistent with earlier measurements. The telescope is operating as
 expected and has met or exceeded all design specifications.

\end{abstract}

\end{frontmatter}

\section{Introduction}

The first detection of Cherenkov light produced by cosmic ray air showers in
the atmosphere was made over half a century ago \cite{Galbraith53}. The
detection method remains essentially unchanged today; relatively crude mirrors
are used to reflect Cherenkov photons onto a photomultiplier tube (PMT)
detector package, and fast electronics discriminate the brief (few nanoseconds)
Cherenkov flashes from the background night-sky light. The first successful
application of this technique to $\gamma$-ray astronomy was realised by the
Whipple Collaboration, who used a $10\U{m}$ diameter light collector and an
array of PMTs to record images of the angular distribution of Cherenkov
light produced by air showers \cite{Weekes77}. The shape and orientation of the images were
used to efficiently select $\gamma$-ray initiated air shower candidate events
from among the otherwise overwhelming background of cosmic ray initiated events
\cite{Hillas85}. The Whipple 10\U{m} telescope provided the first detection of
an astrophysical $\gamma$-ray source, the Crab Nebula, using
this imaging technique \cite{Weekes89}. Arrays of imaging atmospheric Cherenkov telescopes
provide a further increase in sensitivity and in angular and energy resolution,
as demonstrated by the HEGRA experiment \cite{Puhlhofer03}. The stereoscopic
imaging atmospheric Cherenkov technique is now being exploited using large
reflectors by four projects worldwide; MAGIC \cite{Lorenz04} and VERITAS
\cite{Weekes02} in the Northern hemisphere and HESS \cite{Hinton04} and
CANGAROO III \cite{Kubo04} in the South. HESS was the first of these to come
online and recent observations have produced a wealth of new discoveries
\cite{Aharonian05}.

The first stage of the VERITAS project will use four telescopes at a site in
Arizona. Each telescope will consist of a $12\U{m}$ diameter segmented reflector
instrumented with a 499 element PMT imaging camera. The
first of these telescopes has been installed at the base camp of the Whipple
Observatory at Mt. Hopkins (at an altitude of $1275\U{m}$) and saw first light
in February 2005. We begin by describing the various elements of the detector
and assess their technical performance. The calibration and data analysis
procedures are then discussed, and the overall performance of the instrument
is characterised using Monte Carlo simulations and the results from preliminary
observations of TeV $\gamma$-ray sources.

\section{The Telescope}
\begin{figure}[h]
  \begin{center}
    \includegraphics*[height=8cm]{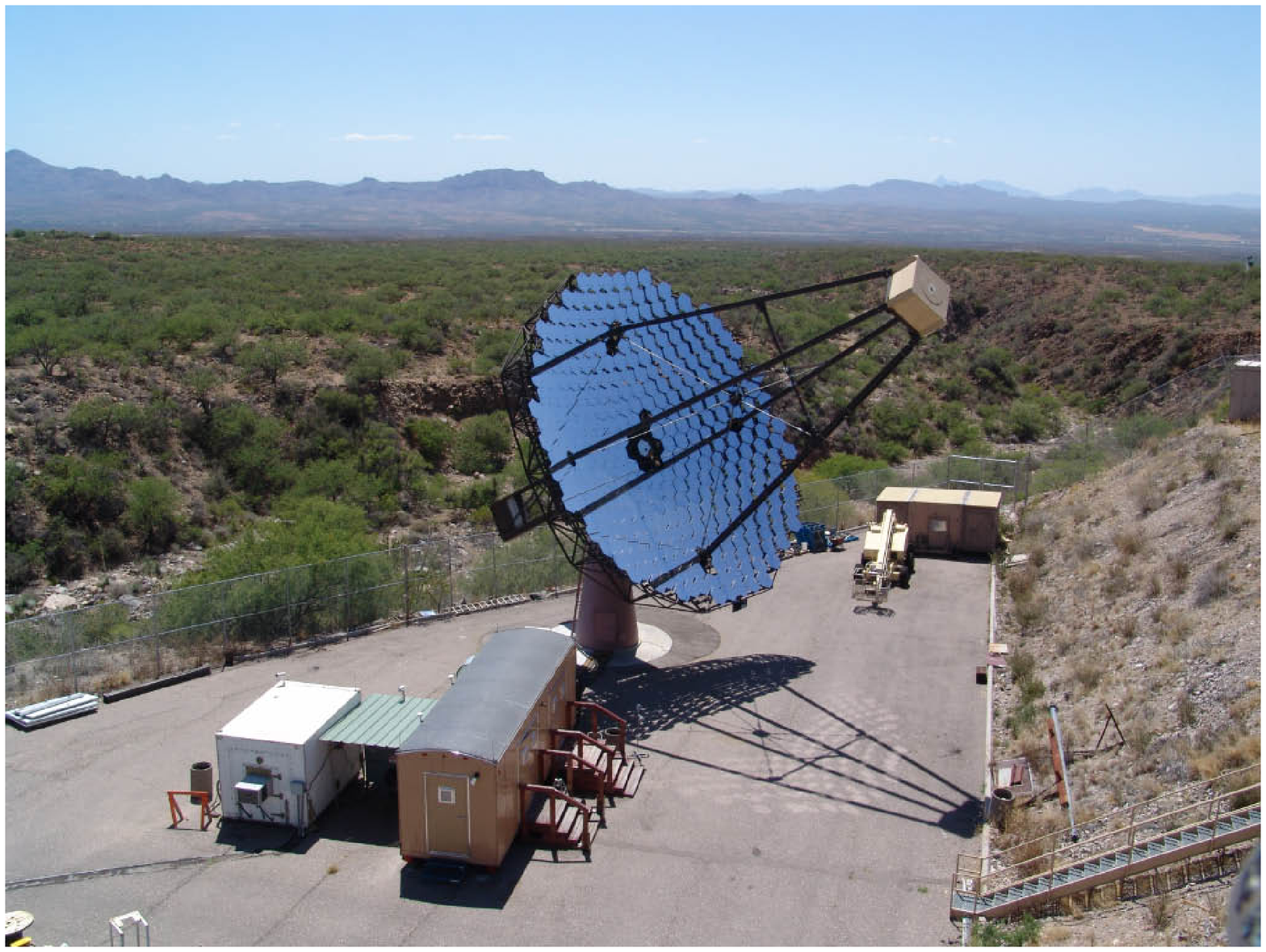}
    \caption{\label{Tel1} 
      The VERITAS Telescope 1 as installed at the Whipple Observatory base camp.
The collector dish has a diameter of $12\U{m}$ and a focal length of $12\U{m}$
and comprises 350 mirror facets. A 499-PMT camera is installed in the
box at the focal point. The buildings in the foreground house the electronics and power supply systems.
    }
  \end{center}
\end{figure}

Figure~\ref{Tel1} shows the first complete VERITAS telescope installed at the
Whipple Observatory base camp. The building closest to the telescope contains
the trigger and data acquisition electronics and is the control room from which
the telescope is operated. The telescope was operated initially in 2004 as a
prototype, with one third of the mirror area and half of the PMTs of the
completed telescope
\cite{Wakely03}.

\subsection{Mechanics and Tracking}
\begin{figure}[h]
  \begin{center}
    \begin{tabular}{cc}
      \includegraphics*[height=7cm]{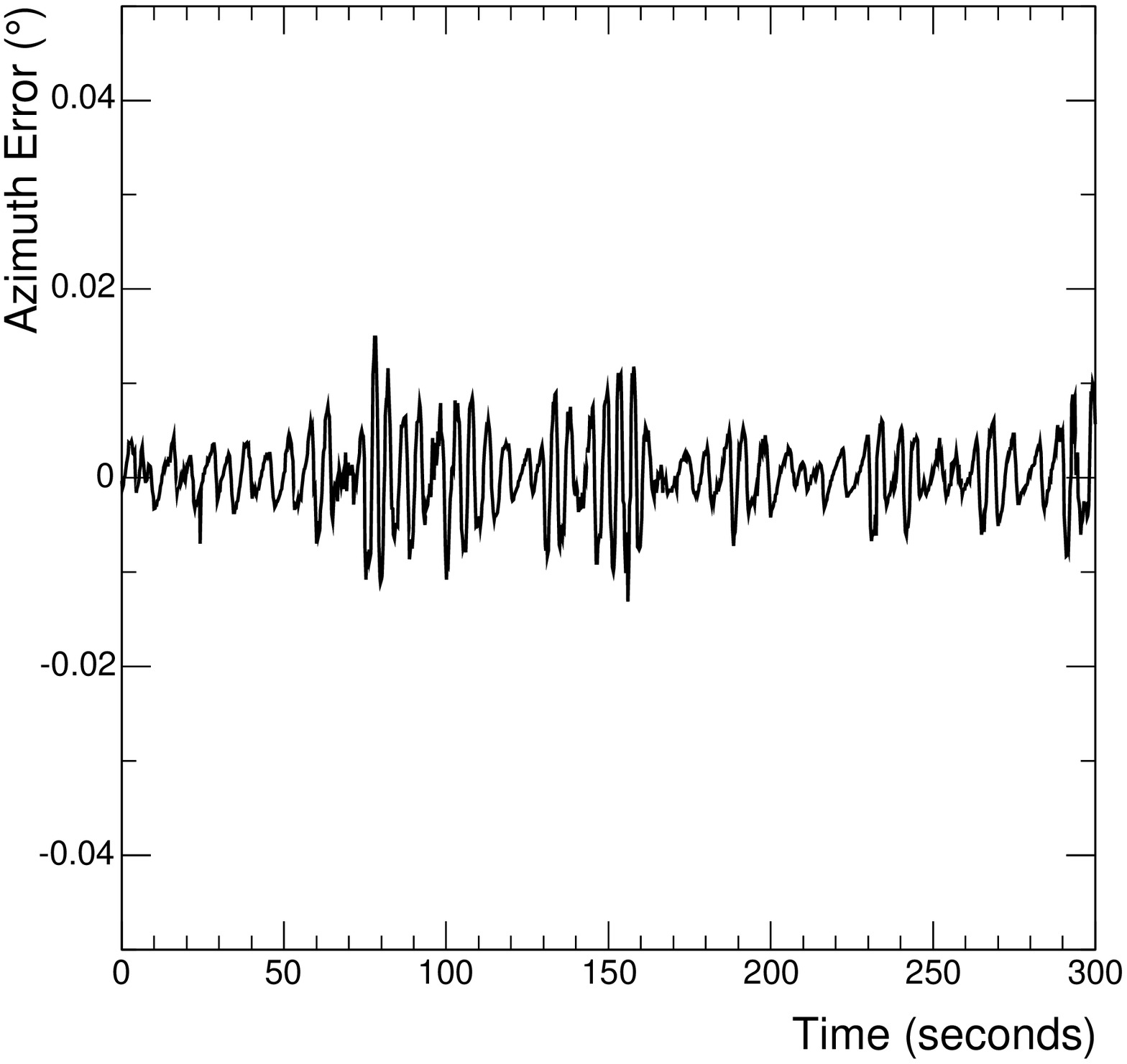}
      \includegraphics*[height=7cm]{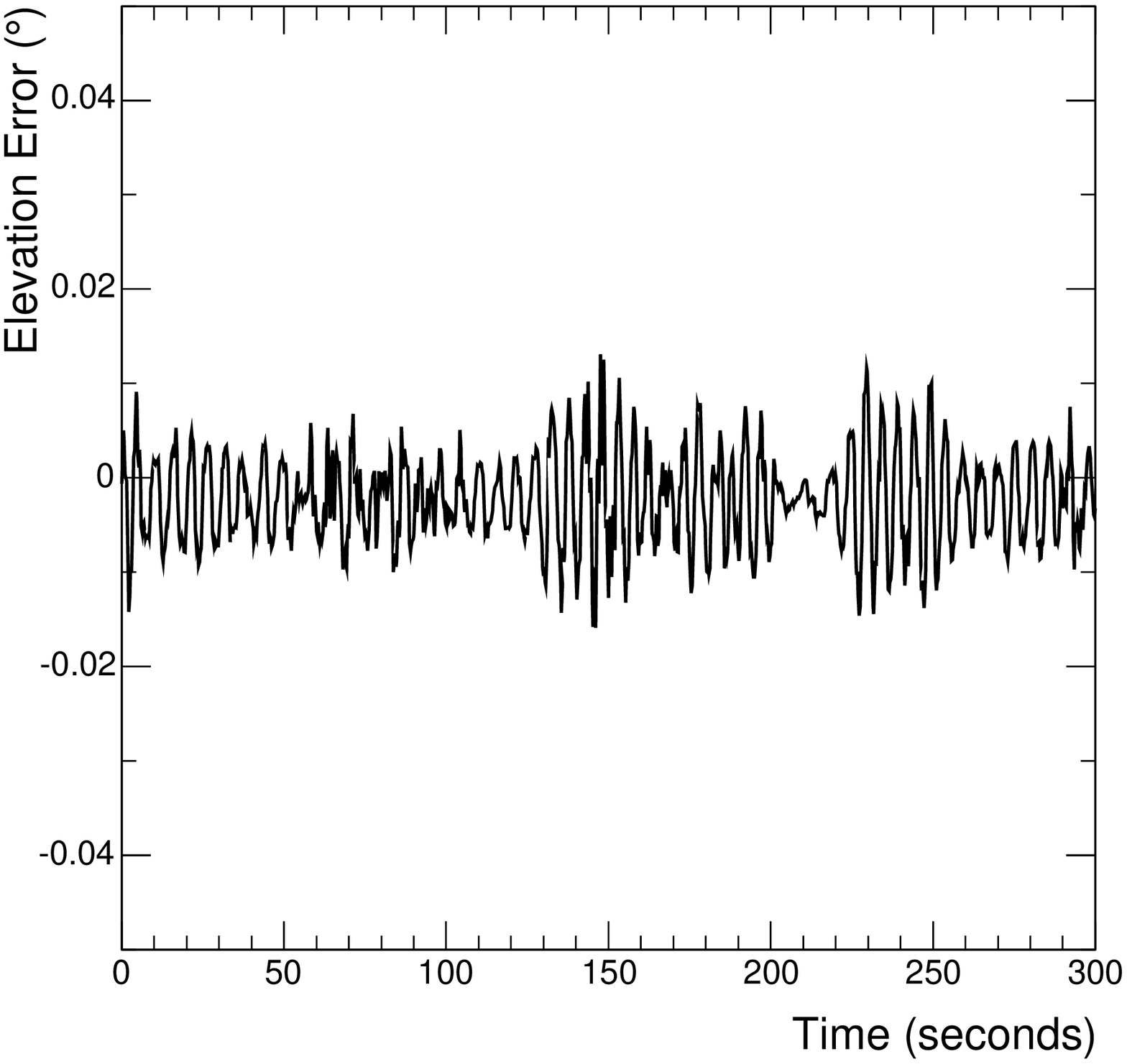}
    \end{tabular}
    \caption{\label{encoders} 
      The azimuth and elevation residuals (difference between measured and
      requested position) for a short tracking run.
    }
  \end{center}
\end{figure}

The basic mechanical structure of the telescope is similar to that of the
Whipple $10\U{m}$ telescope, consisting of an altitude-over-azimuth positioner
and a tubular steel Optical Support Structure (OSS). The camera is supported
on a quadropod, and a mechanical bypass of the upper quadropod arm transfers
this load directly to the counterweight support. The positioner is a
commercial unit manufactured by RPM-PSI (Northridge, California); the OSS is a
steel space frame, custom designed by M3 engineering (Tucson, Arizona) and
fabricated by Amber Steel (Chandler, Arizona) \cite{Gibbs03}.  The maximum
slew speed is measured to be $0.3^{\circ}\UU{s}{-1}$. Tests with a modified
drive system have proved that it is possible to reach maximum slew speeds of
$1^{\circ}\UU{s}{-1}$; the first telescope will shortly be upgraded and the remaining
VERITAS telescopes will have this modification installed as standard. The
telescope encoder measurements are written to a database at a rate of
$4\U{Hz}$ and indicate that the tracking is stable with a relative raw
mechanical pointing accuracy of typically $< \pm0.01^{\circ}$
(Figure~\ref{encoders}). Various methods for measuring and improving the
absolute pointing accuracy are under investigation.

\subsection{Optics}

\begin{figure}[h]
  \begin{center}
    \includegraphics*[height=8cm]{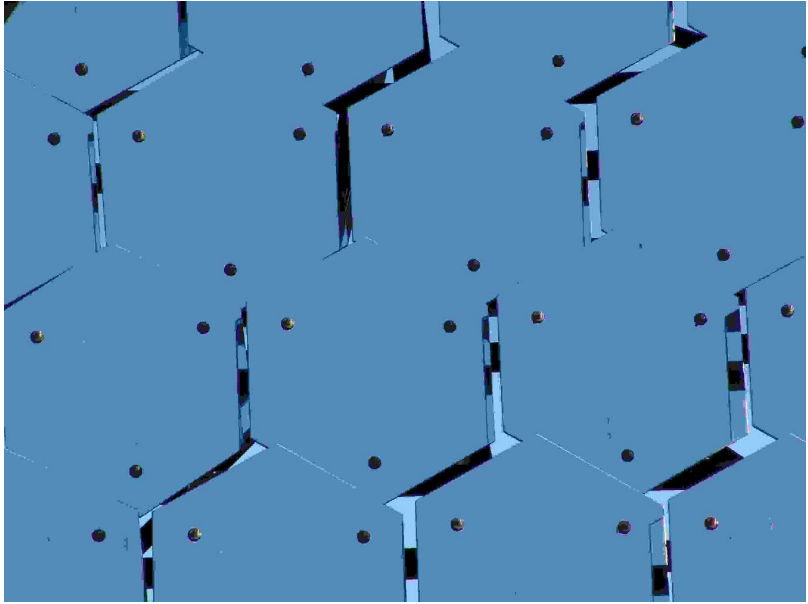}
    \caption{\label{mirrors} 
      A close-up view of the hexagonal VERITAS mirror facets. The three adjustment points which are used in
aligning each facet can be easily seen.
    }
  \end{center}
\end{figure}

\begin{figure}[h]
  \begin{center}
    \includegraphics*[height=8cm]{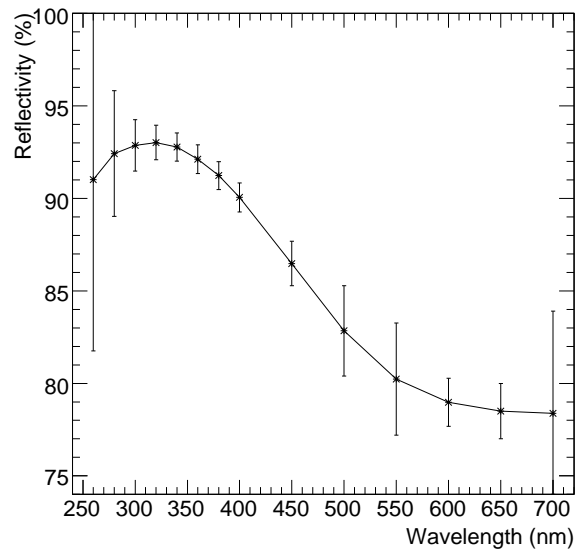}
    \caption{\label{reflect} 
      Average mirror reflectivity as a function of wavelength for photons
      normal to the mirror. Measurements were made in the laboratory after the
      mirrors were produced.
    }
  \end{center}
\end{figure}

The telescope optics follows a Davies-Cotton design \cite{Davies57}, but with
a $12\U{m}$ aperture reflector and a $12\U{m}$ focal length. The reflector
comprises 350 hexagonal mirror facets (Figure~\ref{mirrors}), each with an
area of $0.322\UU{m}{2}$, giving a total mirror area of
$\sim110\UU{m}{2}$. The use of hexagonal facets allows the full area of the
OSS to be exploited. The facets are made from glass, slumped and polished by
DOTI (Roundrock, Texas), then cleaned, aluminized and anodized at the VERITAS
optical coating laboratory. The reflectivity of the anodized coating at normal
incidence is shown as a function of wavelength in Figure~\ref{reflect}; it is
typically $>90\U{\%}$ at $320\U{nm}$. Each facet has a $24.0\U{m}\pm1\U{\%}$
radius of curvature and is mounted on the spherical front surface of the OSS
(radius $12\U{m}$) using a triangular frame. Three adjustment screws allow
each facet to be accurately aligned.

\begin{figure}[h]
  \begin{center}
    \includegraphics*[height=8cm]{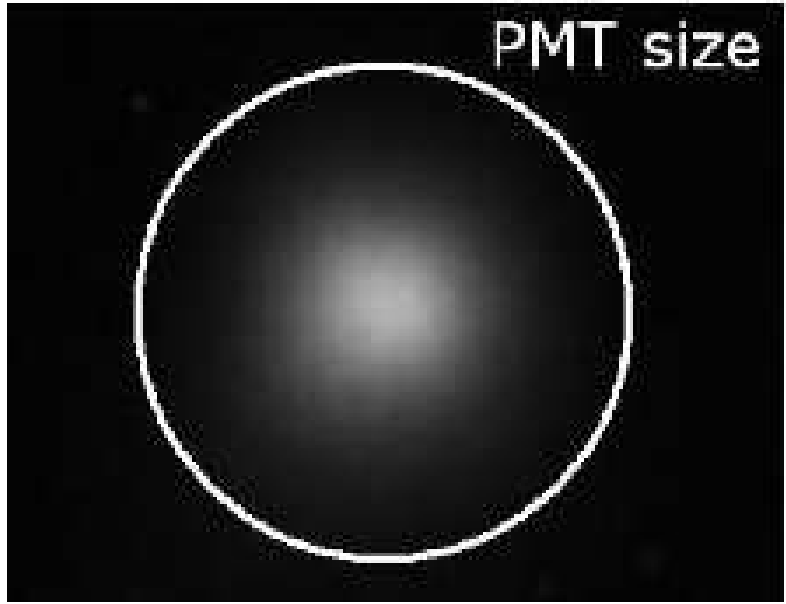}    
    \caption{\label{PSF} 
      An image of Polaris in the focal plane of the telescope recorded with a
      CCD camera. The point spread function is $0.06^{\circ}$ FWHM. The circle
      indicates the size of a VERITAS PMT ($0.15^{\circ}$ diameter).
    }
  \end{center}
\end{figure}

The reflector facets are aligned manually using a laser system installed at a
point facing the centre of the reflector at a distance of twice the focal
length ($24\U{m}$). An initial alignment of the mirror facets resulted in a
point spread function (PSF) of $\sim0.09^{\circ}$ FWHM. The $\gamma$-ray
observations reported on in this paper were made under this alignment
condition. A subsequent realignment reduced the PSF to $0.06^{\circ}$
(Figure~\ref{PSF}) at the position of Polaris (elevation $31^{\circ}$),
degrading at higher elevations due to flexure of the OSS. The technique of
bias alignment, wherein the mirror facets are aligned such that the PSF is
optimum over the most useful observing range, has been successfully employed
on the Whipple $10\U{m}$ telescope in the past and will be applied to the
VERITAS telescopes to achieve a PSF of $<0.06^{\circ}$ FWHM over the
$40^{\circ}-80^{\circ}$ elevation range.

\subsection{Camera}
\begin{figure}[h]
  \begin{center}
    \includegraphics*[height=8cm]{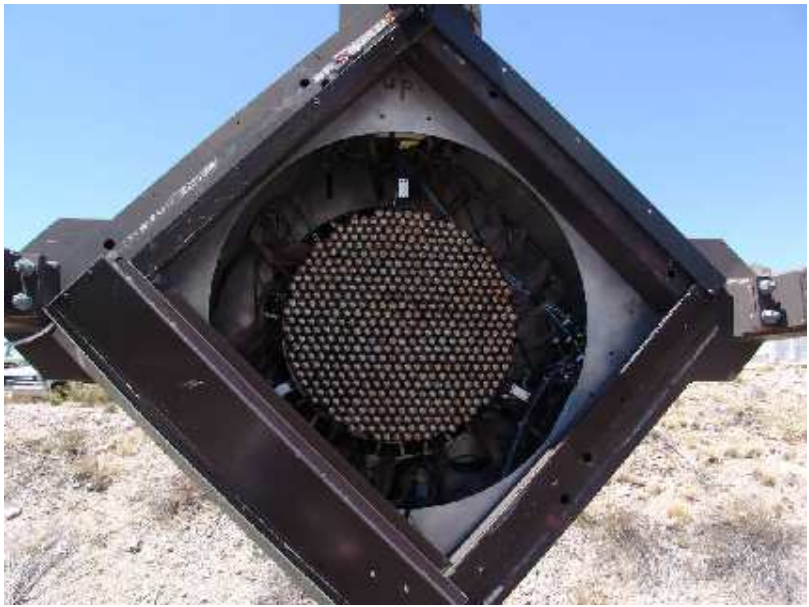}
    \caption{\label{camera} 
      The 499 PMT pixel camera. The focus box is $1.8\U{m}$ square. A remotely
      operated shutter usually covers the camera during daylight hours.  
    }
  \end{center}
\end{figure}

Figure~\ref{camera} shows the front face of the telescope focus box. The box
is $1.8\U{m}$ square and allows for future expansion to increase the camera
field-of-view. The imaging camera's 499 pixels are Photonis $2.86\U{cm}$
diameter, UV sensitive PMTs (model XP2970/02), with a quantum efficiency
$>20\%$ at $300\U{nm}$. The angular pixel spacing is $0.15^{\circ}$, giving a
total field-of-view of diameter $\sim3.5^{\circ}$. Light cones have not yet
been installed; two different designs are being fabricated and will be
tested. They are expected to significantly improve the total photon collection
efficiency as well as to reduce the rate of off-axis background photons in
each PMT.

The PMT high voltage is provided by a multichannel modular commercial power
supply (CAEN) which allows each PMT to be controlled individually. The high
voltage is chosen to give a PMT gain of $\sim2\times10^5$. The signals are amplified by a
high-bandwidth pre-amplifier integrated into the PMT base mounts. This circuit
also allows the PMT anode currents to be monitored and calibrated charge
pulses to be injected into the signal chain. Average currents are typically
$3\U{\mu A}$ (for dark fields) to $6\U{\mu A}$ (for bright fields),
corresponding to a night-sky photoelectron background of 100-200\U{MHz}
per PMT at this site. Signals are sent via $\sim50\U{m}$ of $75\U{\Omega}$
stranded cable (RG59) to the telescope trigger and data acquisition
electronics which is housed in the control room.

\subsection{Trigger}
\label{trigtext}
In an imaging atmospheric Cherenkov telescope, precise timing between trigger
channels is desirable in order to reduce the coincidence resolving time and
hence lower the detector energy threshold. To achieve this, each channel is
equipped with a custom-designed constant fraction discriminator (CFD)
\cite{Hall03} which has a trigger time which is independent of the input pulse
height. The CFD output width is programmable in 12 steps between 4 and
$25\U{ns}$; a width of $10\U{ns}$ was used as standard for telescope
operations in 2005. A 3-bit, $6\U{ns}$ programmable delay is provided for each
channel so as to correct for systematic differences in the relative signal
paths due to cable length differences and the voltage-dependent PMT transit
times.

\begin{figure}[h]
  \begin{center}
    \includegraphics*[height=8cm]{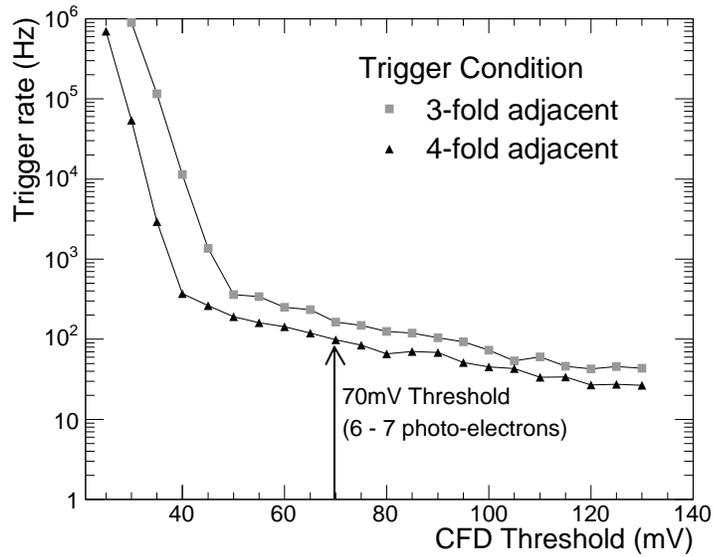}
    \caption{\label{bias} 
     The trigger rate as a function of CFD threshold for two different
     topological trigger configurations. 
    }
  \end{center}
\end{figure}

The CFD signals are copied and sent to a topological trigger system which is
similar to that used successfully on the Whipple $10\U{m}$ telescope
\cite{Bradbury02}, but with an improved channel-to-channel timing jitter of
$<1\U{ns}$. The system contains a memory look-up which can be pre-programmed
in a few minutes to recognise patterns of triggered pixels in the camera; for
example, any 3 adjacent pixels. The required overlap time between adjacent CFD
signals is $\sim6\U{ns}$. The topological trigger system reduces the rate of
triggers due to random fluctuations of the night-sky background light and
preferentially selects compact Cherenkov light images. Figure~\ref{bias}
shows the trigger rate as a function of CFD threshold for two different
topological trigger configurations. Observations in 2005 were all made with a
rather conservative CFD threshold corresponding to $\sim6-7$ photoelectrons
and a 3-fold adjacent pixel topological trigger configuration, giving a cosmic
ray trigger rate at high elevation of $\sim150\U{Hz}$.

VERITAS will also include an array level trigger system which triggers the data
acquisition on coincident events over a pre-defined number of telescopes. A
preliminary array level trigger is installed on the first VERITAS telescope
which is used to latch the event number and GPS time for each event.

\subsection{Data Acquisition}

\begin{figure}[h]
  \begin{center}
    \includegraphics*[height=8cm]{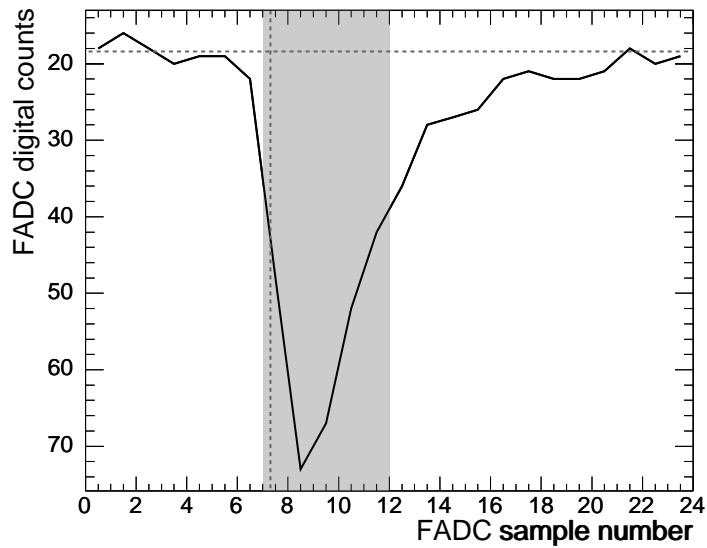}
    \caption{\label{trace} A FADC trace produced by Cherenkov light from a
      cosmic ray air shower in a single signal channel. The dashed horizontal
      line is the electronic pedestal level, the dashed vertical line shows
      the pulse arrival time, $t_{0}$, and the shaded area indicates a
      $10\U{ns}$ (5 FADC samples) integration window.}
  \end{center}
\end{figure}

The centrepiece of the data acquisition chain is a custom-built $500\U{MHz}$
flash-ADC system \cite{Buckley03}. Each PMT signal is digitized with an 8-bit
dynamic range and a memory depth of $32\U{\mu s}$. By default, the signal
traces follow a high gain path to the FADC. If the dynamic range is
exceeded, an analog switch connects the FADC chip to a delayed low gain
path instead, thus extending the dynamic range for each $2\U{ns}$
sample. The electronic noise is small, with a sample-to-sample standard deviation of
$\sim0.5\U{digital~counts}$ and an event-to-event standard deviation over a 10 sample integration
window of $\sim1.5\U{digital~counts}$ Figure~\ref{trace} shows a typical FADC
trace in a signal channel.

The FADCs are deployed in four custom VME crates, where they are read out by
local single board computers (VMIC). Buffered events from the single board
computers are transferred via a Scaleable Coherent Interface to an
event-building computer, where they are integrated, tested, and passed on to
the online analysis system and data harvester. The FADC readout window size
and position are programmable; a 24 sample readout on all 500 channels results
in a data size of $13.5\U{kb}$ per event and a deadtime of $\sim10\U{\%}$ at
$150\U{Hz}$. While this is manageable for a single telescope, the VERITAS
array will produce four times as much data at higher rates (up to
$1\U{kHz}$). To cope with this, a zero suppression scheme has been implemented
wherein only those channels with a peak signal larger than some preset value
are read out, reducing the data size by a factor of $>4$. The CFDs are
integrated onto the FADC boards and the CFD hit pattern and the rate of single
channel triggers can also be read out to the data stream.

The FADCs give two principal benefits over simple charge integrating ADCs.
They allow the application of digital signal processing techniques, for
example, actively placing and minimizing the charge integration gate, thus
improving the signal/noise per PMT and lowering the effective energy threshold
in the analysis. In addition, they provide measurements of the time
distribution of the Cherenkov photons across the image, which may help to
reject the hadronic cosmic ray background showers and improve the accuracy of
$\gamma$-ray shower parameter reconstruction \cite{Holder05}.

\section{Calibration}
The telescope calibration is divided into two sections. Absolute calibration
is concerned with understanding the signal size produced by a single photon
such that detector response and the energy scale can be accurately modelled in
Monte Carlo simulations. Relative calibration involves measuring the various
calibration constants for each signal channel so as to flat-field the response
of the camera in the data analysis.

\subsection{Absolute Calibration}
\label{abs_calib}
\begin{figure}[h]
  \begin{center}
    \includegraphics*[height=8cm]{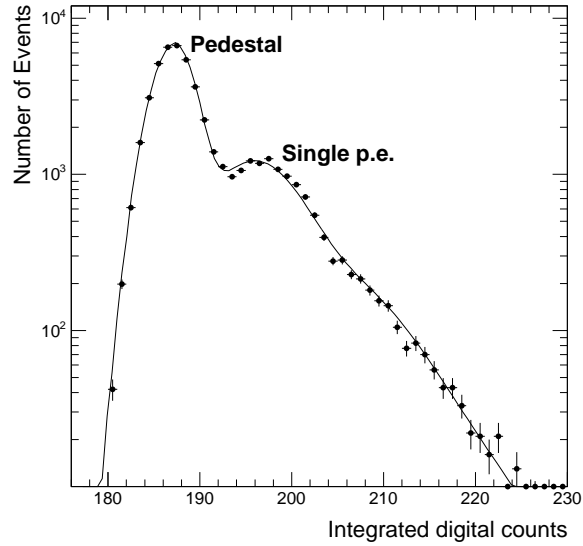}
    \caption{\label{singlepe}
      The single photoelectron response for one PMT at standard operating
      gain. The fit assumes a Poisson distribution of photoelectrons and a
      Gaussian distribution for the integrated charge produced by a single
      photoelectron. The average conversion factor over all PMTs is $0.19\pm0.02$ photoelectrons per digital count.
    }
  \end{center}
\end{figure}

The overall photon conversion factor of the telescope is a combination of the
mirror reflectivity, the collection efficiency and quantum efficiency of the
PMT photocathode, and the conversion factor of the electronics chain. The mirror
reflectivity and PMT photocathode characteristics can be determined from laboratory
measurements and manufacturer's specifications; the conversion factor of the electronics
chain is defined as the average integral number of FADC counts produced by the
generation of a single photoelectron in the PMTs and measured \textit{in
situ}. A nitrogen laser is used to deliver a $\sim4\U{ns}$ wide pulse to a dye
module which fluoresces at $400\U{nm}$. This light is passed via optical fibre
to an opal glass diffuser situated in front of the camera at a distance of
$\sim4\U{m}$, so as to provide uniform illumination \cite{Hanna02}. The camera face is
covered by a semi-reflective mylar sheet which shields the PMTs from background
light, and the laser pulse intensity is adjusted using filters to provide an
average illumination of $\sim1\U{photoelectron}$ per PMT. A delayed copy of the laser
trigger is used to trigger the data acquisition such that the signal arrives
centered in the readout window.
Figure~\ref{singlepe} shows the histogram of the integrated FADC counts for a
single PMT at standard operating gain. The conversion factor averaged over all
PMTs is $0.19\pm0.02$ photoelectrons per digital count.

Two complementary methods have been used to confirm the overall photon
collection efficiency of the telescope. Single muons within the field-of-view
produce sharply defined ring images, and the expected amount of Cherenkov
light per unit arc length can be accurately calculated, providing a well
calibrated natural ``test beam'' \cite{Vacanti94}. A comparison of the total
charge in simulated muons and in real muons selected from data taken with the
first VERITAS telescope indicates that the total photon collection efficiency
is well matched \cite{Humensky05}. A novel alternative approach to absolute
calibration, borrowed from atmospheric fluorescence detectors of ultra-high
energy cosmic rays \cite{Roberts03}, has also been tested. A calibrated laser
pulse directed towards the zenith is used to produce a flash of
Rayleigh-scattered light with an intensity that can be calculated very
accurately when atmospheric conditions are good. Preliminary results with this
method again indicate good agreement between simulations and data
\cite{Shepherd05}.

\subsection{Relative Calibration}
Prior to the parameterization of Cherenkov images in the camera, each signal
channel must be calibrated. A pulse generator is used to trigger the data
acquisition at a rate of $3\U{Hz}$ during the data-taking in order to generate
events with no Cherenkov light present. A histogram is constructed of the
integrated number of digital counts in each FADC trace (for a given integration window size);
the mean of this histogram is the electronic pedestal value and the width is
the pedestal standard deviation, which provides a measure of the night sky background
noise level. Pedestal and pedestal standard deviation values are currently calculated
once per 28 minute data run. The same laser system as used for the single
photoelectron calibration is used once per night to provide $\sim1000$ bright,
uniform, time-coincident photon pulses across the camera. The mean of a
histogram of the integrated signal in each FADC trace over these events
measures the relative gain of the signal channels. Similarly, a histogram of
the arrival time of the laser pulse in each trace measures the relative time
offsets for each channel. Relative PMT gains are set to within $\sim10\%$ by
adjusting the PMT high voltages. Approximately 95\% of the signal channels
show a relative time offset of less than $\pm3\U{ns}$; this is compensated for in
the trigger hardware as discussed in Section~\ref{trigtext}.

\section{Data Analysis and Results}
\label{analysis}
Development work to define the optimum data analysis methods for this new
instrument is ongoing. The data products, in particular the FADC information,
provide the opportunity of exploiting many techniques which have
not been applicable to previous generations of Cherenkov telescopes; for
example, the use of digital signal processing algorithms to parameterise the
signal traces, and the added dimension of detailed image timing
information. The analysis described here is based on that which has been
successfully employed for the analysis of Whipple $10\U{m}$ telescope data for many
years, with the use of image timing information being restricted to helping to
define the optimum signal integration window.

\begin{figure}[h]
  \begin{center}
    \includegraphics*[height=8cm]{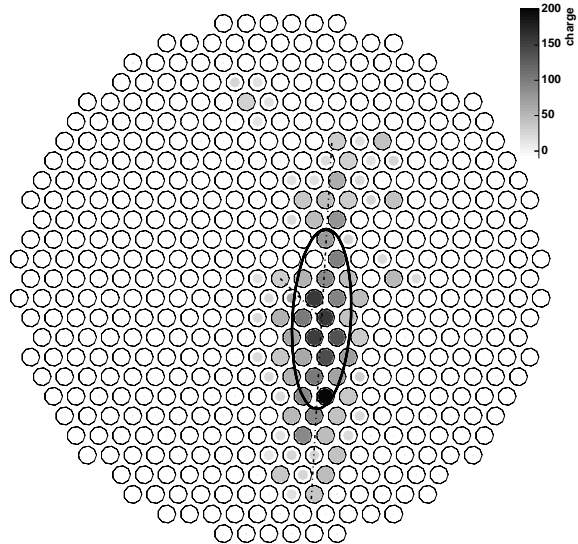}
    \caption{\label{event}
      The charge distribution across the camera for a cosmic ray
      event (the grey scale is in digital counts).
    }
  \end{center}
\end{figure}

\begin{figure}[h]
  \begin{center}
    \includegraphics*[height=8cm]{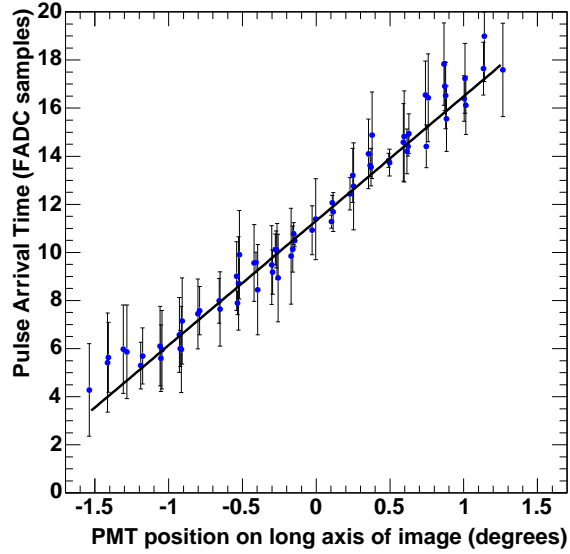}
    \caption{\label{tgrad}
      The Cherenkov pulse arrival time
      distribution (in units of FADC samples = $2\U{ns}$) along the long axis
      for the cosmic ray image shown in Figure~\ref{event}. A PMT position of
      zero corresponds to the centre of the image ellipse. Error bars
      correspond to the signal size-dependent time resolution for each PMT,
      as measured using the laser calibration system \cite{Holder05}.
    }
  \end{center}
\end{figure}

We have used a two-pass method in this analysis. In the first pass, a wide
($20\U{ns}$) integration gate is applied to each FADC trace at a fixed time
position in order to calculate the integrated charge and the pulse arrival
time, $T_0$ (defined as the half-height point on the pulse leading edge). PMTs
are selected which produce an integrated signal greater than $5$ times their
pedestal standard deviation, or greater than $2.5$ times the pedestal standard
deviation while also being adjacent to a PMT selected at the higher
threshold. The resulting shower image is then parameterised with a second
moment analysis, the results of which can be described by an ellipse as shown
in Figure~\ref{event} \cite{Hillas85}. Figure \ref{tgrad} shows $T_0$ as a
function of PMT position along the long axis of the ellipse for this event;
the time gradient across the image is given by a straight line fit to this
graph. In the second pass over the image, a shorter ($10\U{ns}$) integration
window is applied to each trace, but the location of the window is calculated
using the expected position of the pulse according to the measured time
gradient across the image.

\begin{figure}[h]
  \begin{center}
    \begin{tabular}{cc}
      \includegraphics*[height=8cm]{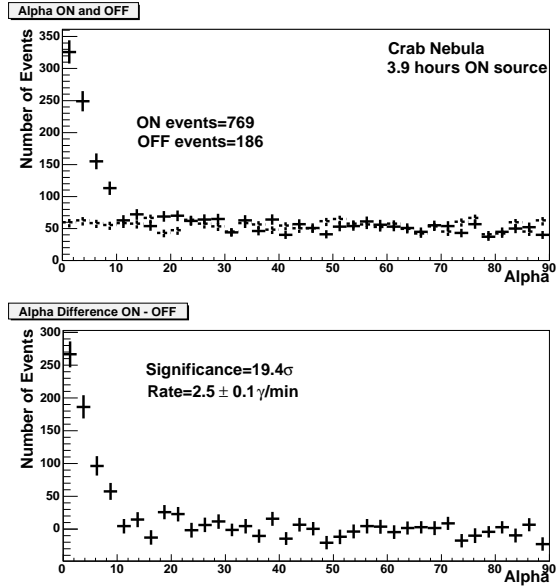}
    \end{tabular}
    \caption{\label {crab}
      The image orientation angle, $\alpha$, as described in the text, for ON and OFF-source
      observations of the Crab Nebula after $\gamma$-ray selection cuts. The
      excess events at low values of $\alpha$ are due to $\gamma$-rays
      originating from the direction of the Crab Nebula. Selecting events with
      $\alpha<7^{\circ}$ results in a detection with a significance of $19.4\sigma$.
    }
  \end{center}
\end{figure}

\begin{figure}[h]
  \begin{center}
    \begin{tabular}{cc}
        \includegraphics*[height=8cm]{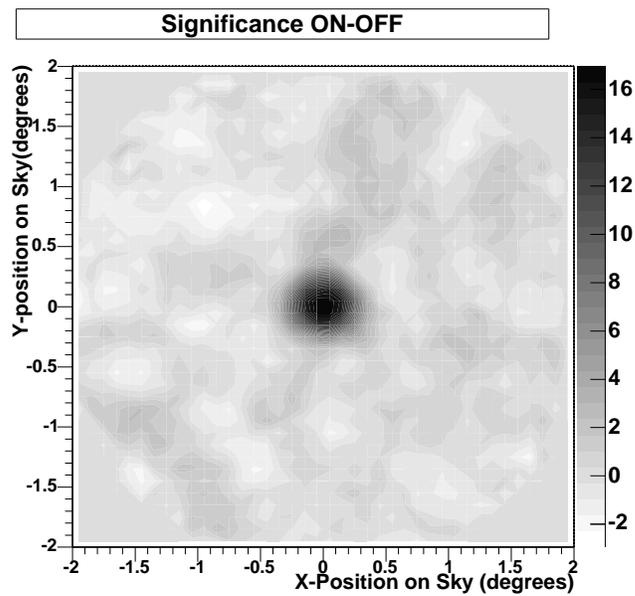}
    \end{tabular}
    \caption{\label {crab2d}
      The two-dimensional map of reconstructed source position in degrees on
      the camera plane (the camera is 3.5$^{\circ}$ in diameter). The Crab
      Nebula was located at the centre of the camera. The angular resolution
      is $0.17^{\circ}$ (68\% containment radius). Note that
    adjacent bins are not statistically independent.
    }
  \end{center}
\end{figure}

The standard candle of ground-based $\gamma$-ray astronomy is the Crab Nebula,
and the sensitivity of any new instrument is best tested against this
benchmark. The first VERITAS telescope saw first light in February 2005,
leaving only a small time window in which to collect data on the Crab while
the source was still observable at high elevations.  Observations were taken
in an ON/OFF mode, wherein a 28 minute ON exposure at the source position is
followed or preceded by a 28 minute OFF-source exposure, at the same elevation
but offset by $\pm30$ minutes in right ascension. The total ON-source exposure
was $3.9\U{hours}$ with the same amount of OFF-source data collected in order
to provide a measure of the cosmic ray background. Due to the purely
electromagnetic nature of $\gamma$-ray initiated air showers, $\gamma$-ray
Cherenkov images are expected to be more compact than cosmic ray images and
can be preferentially selected by applying cuts based on the dimensions of the
event ellipse ($length$, $width$) \cite{Hillas85}. The parameter $\alpha$
describes the orientation of the long axis of the image ellipse relative to
the line joining the centre of the field-of-view and the ellipse
centroid. Showers originating from a point source, such as the Crab Nebula,
will be oriented with their long axes pointing back to the position of the
source at the centre of the camera. The angular $distance$ of the image from
from the source position is related to the distance of the shower core
position on the ground. Finally, the ratio of the image $length$ to its total
$size$ (where $size$ is the integrated charge over all PMT signals in the
image) is used to discriminate $\gamma$-rays from the otherwise overwhelming
background of local muons which generate short arcs in the camera with a
constant size per unit arc length. Cuts on these various parameters were
optimised using a data set of observations of the variable TeV blazar
Markarian 421 when this source was known, from contemporaneous observations
with the Whipple~10m telescope, to be in a high emission state. The resulting
$\alpha$ histogram for the Crab Nebula observations is shown in
Figure~\ref{crab}. The excess at low $\alpha$ corresponds to a significance of
$19.4\sigma$ (where $\sigma$ refers to one standard deviation above the
background), indicating a sensitivity of $\sim10\sigma$ for one hour of
ON-source observations. Figure~\ref{crab2d} shows the significance map of the
reconstructed source position using the method of Lessard et
al. \cite{Lessard01}. 


Following the Crab Nebula observations, the telescope was used in an observing
campaign throughout 2005, complementary to that of the Whipple 10~m telescope,
which resulted in $>10\sigma$ detections of the known $\gamma$-ray sources
Markarian~421 and Markarian~501 \cite{Cogan05} and data sets on various
potential TeV sources. A full online analysis package has been developed and
was successfully used to detect strong $\gamma$-ray flaring behaviour from
Markarian 421 on a timescale of minutes.

\section{Simulations}
\subsection{The Simulation Chain}

In order to accurately calculate source fluxes and energy spectra, it is
necessary to develop a detailed model of the telescope performance, to which
simulated air showers are presented such that the telescope detection
efficiencies can be calculated. A complete chain of Monte Carlo simulations
has been developed \cite{Maier05}.  It consists of air shower simulations with
CORSIKA \cite{Heck} and a detailed simulation of the telescope response (GrISU
\cite{LeBohec}).  CORSIKA version 6.20 is used with the hadronic interaction
models QGSJet for primary energies above $500\U{GeV}$ and with FLUKA for
primary energies below $500\U{GeV}$.  Simulated hydrogen, helium, and oxygen
nuclei-induced air showers were produced in an energy range from $10\U{GeV}$
($150\U{GeV}$ for oxygen) to $50\U{TeV}$ for various telescope elevation
angles.  Spectral indices were taken from fits to balloon measurements
\cite{Hoerandel}.  The shower cores were chosen to be distributed randomly
within a circle of radius $1000\U{m}$ centred on the telescope and the
isotropic distribution of the cosmic ray incidence angles was simulated by
randomizing the shower directions in a cone of radius 3.5$^{\circ}$ around the
pointing direction of the telescope (while showers from outside of this cone
angle can trigger the telescope, we estimate the fraction to be no more than
5\% of the total).  Primary $\gamma$-rays were simulated
with a Crab-like spectrum ($E^{-2.6}$) in an energy range from $10\U{GeV}$ to
$10\U{TeV}$ on a circular area at the ground with radius $450\U{m}$.

Measurements of the atmospheric properties at the site of the first
VERITAS telescope are currently in progress.  The calculations described here
used the U.S. standard atmosphere, which does not always reflect the
properties of the atmosphere in Southern Arizona.  Atmospheric extinction
values were estimated using MODTRAN 4 \cite{Modtran} assuming $50\U{km}$
visibility at  $550\U{nm}$ at ground level.  The photoelectron rate per PMT
is measured to be 100-$200\U{MHz}$, corresponding to a  night sky background
rate of $\sim2.8\times 10^{12}$ photons m$^{-2}$ s$^{-1}$ sr$^{-1}$,
which was used in the simulations and produced a simulated pedestal
standard deviation of a similar level to that observed in the data.

The telescope simulations consist of two parts, the propagation of Cherenkov
photons through the optical system and the response of the camera and
electronics.  The geometrical properties of the optical system are fully
implemented in the simulation, including allowance for the surface roughness
of mirror facets and for random scatter in their alignment.  The camera
configuration corresponds to that of April 2005, i.e.~a 499 pixel camera with
$2.86\U{cm}$ diameter phototubes without light cones, with a total
field-of-view of 3.5$^{\circ}$.  The absolute gain calibration has been
described in Section~\ref{abs_calib}. In order to estimate the exact pulse
shape to be used in the simulations, we examined pulses in the data produced
by muon events and laser flashes, both of which are light sources with a time
profile which is expected to be shorter than the bandwidth of the electronics.
The resulting average single photoelectron pulse has a rise time of $3.3\U{ns}$ and
a width of $6.5\U{ns}$. Cherenkov photons hitting a PMT produce a single
photoelectron pulse with appropriate amplitude and time jitters applied.
Electronic noise and all efficiencies, including mirror reflectivities,
geometrical, quantum, and collection efficiencies, and losses due to signal
transmission were modeled.  The pulses are digitized into $2\U{ns}$ samples with a
trace length of 24 samples reflecting the properties of the FADC system.  The
trigger simulation uses a simplified model of the constant fraction
discriminator, which is the first stage of the VERITAS multi-level trigger,
and a full realization of the pattern trigger, requiring three adjacent pixels
above threshold in a time window of 6 ns.  The currently used trigger
threshold of 70 mV corresponds to about 6.7 photoelectrons.  The output of the
telescope simulations, i.e.~FADC traces for all PMTs, are written to disk in
the VERITAS raw data format.

\subsection{Raw Trigger Rate}
The raw trigger rate of the telescope with trigger conditions as described
above is $\sim150\U{Hz}$ at high elevation.  Dead time losses due to readout are
$\sim$10\%, the corrected trigger rate is consequentially $\sim$160 Hz.  The
simulation of the cosmic ray background results in trigger rates of 101 Hz
from air showers induced by protons,  26 Hz from helium nuclei, and $\sim$5 Hz
from nuclei of the CNO group for an elevation of 70$^{\circ}$.  The Monte
Carlo calculations reproduce the observed trigger rate with an accuracy of
$\sim$20\%, which is acceptable, taking into account an estimated uncertainty
in the cosmic ray fluxes of $\sim$25\%  and various systematic uncertainties
in the modeling of the telescope and the atmosphere.

\subsection{Comparison with Real Data}

Simulations and real data were analysed in the same way. The analysis method
used was a simpler, single-pass version of the analysis described in
Section~\ref{analysis}, which has a slightly reduced sensitivity.
$\gamma$-ray candidates were extracted from observations of the Crab Nebula
(3.9 hours ON-source) and Markarian 421 (4.2 hours ON-source) in March and
April 2005.  All data were taken in ON/OFF mode at elevations above
60$^{\circ}$ and in good weather conditions. The $\gamma$-ray selection cuts
result in a combined significance of 21.7$\sigma$ and 900 $\gamma$-ray
candidates for the combined observations of the Crab Nebula and Mrk 421. The
measured $\gamma$-ray rate from the observations of the Crab Nebula alone is
$2.1\pm0.2 \UU{minute}{-1}$. Using the Crab spectrum reported in
 Hillas et al. \cite{Hillas98} results in a simulated $\gamma$-ray rate from the Crab Nebula of
$2.2\UU{minute}{-1}$.

\begin{figure}[ht]
\begin{center}
\includegraphics*[width=0.41\textwidth]{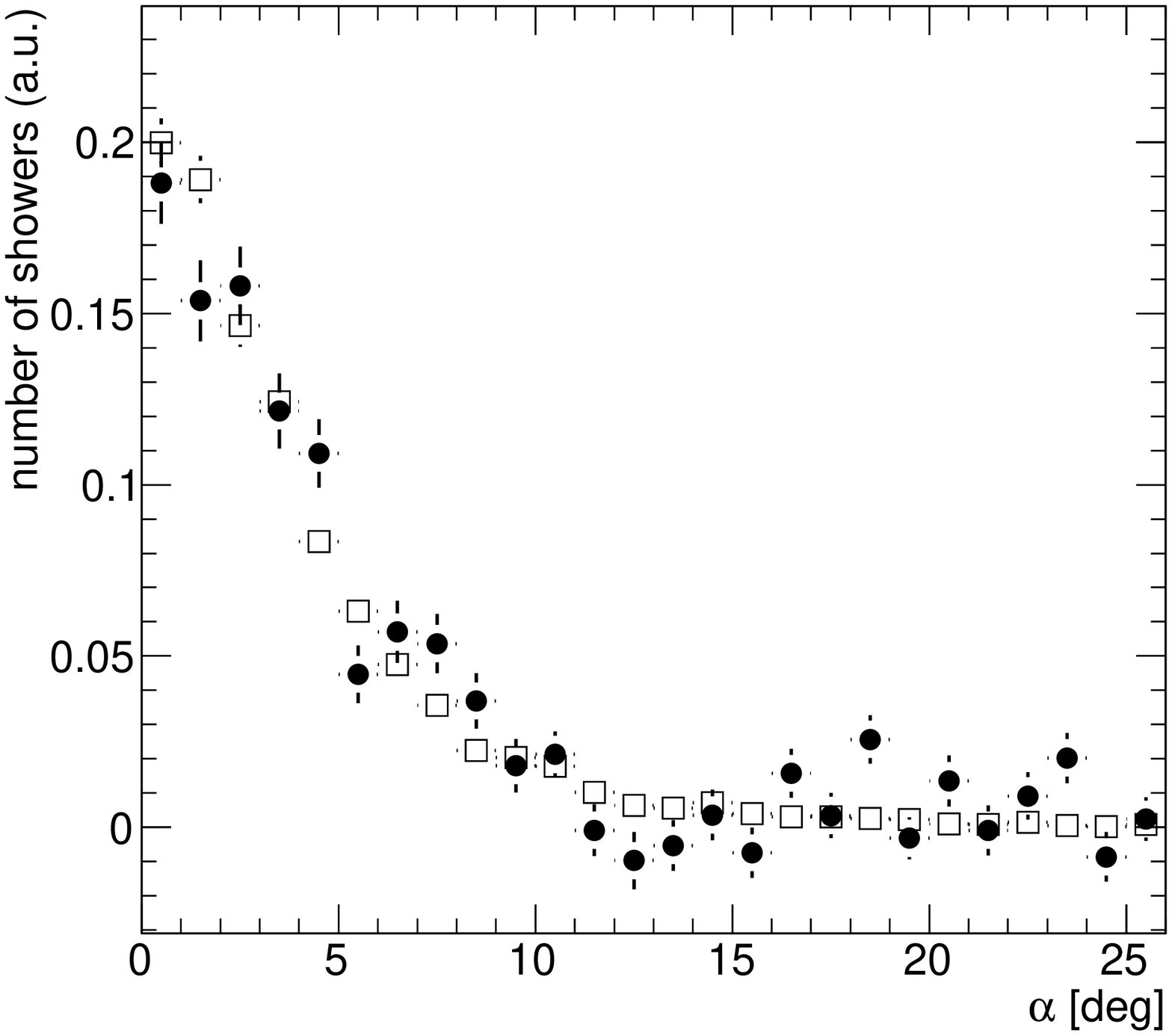}
\includegraphics*[width=0.41\textwidth]{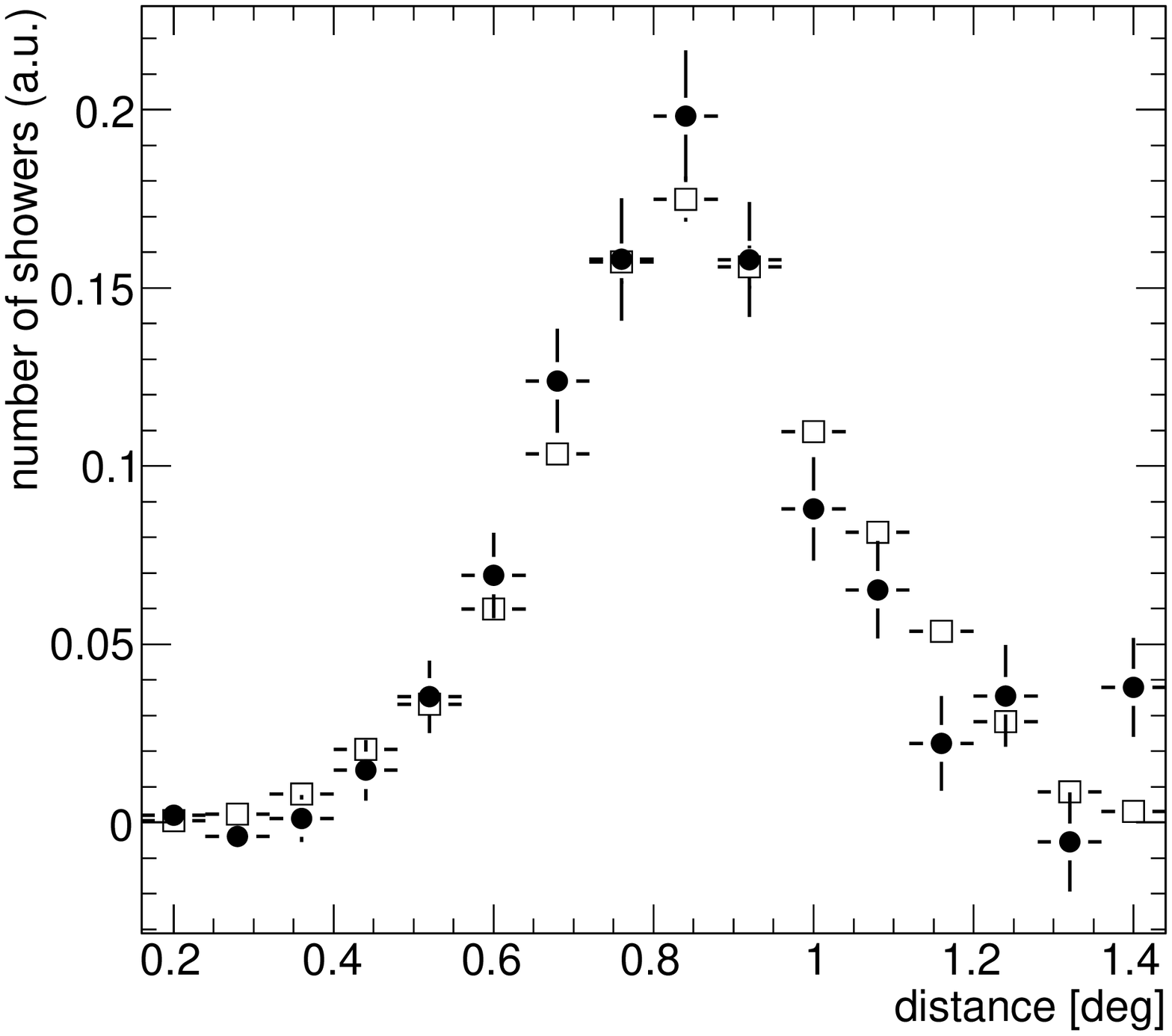}
\includegraphics*[width=0.41\textwidth]{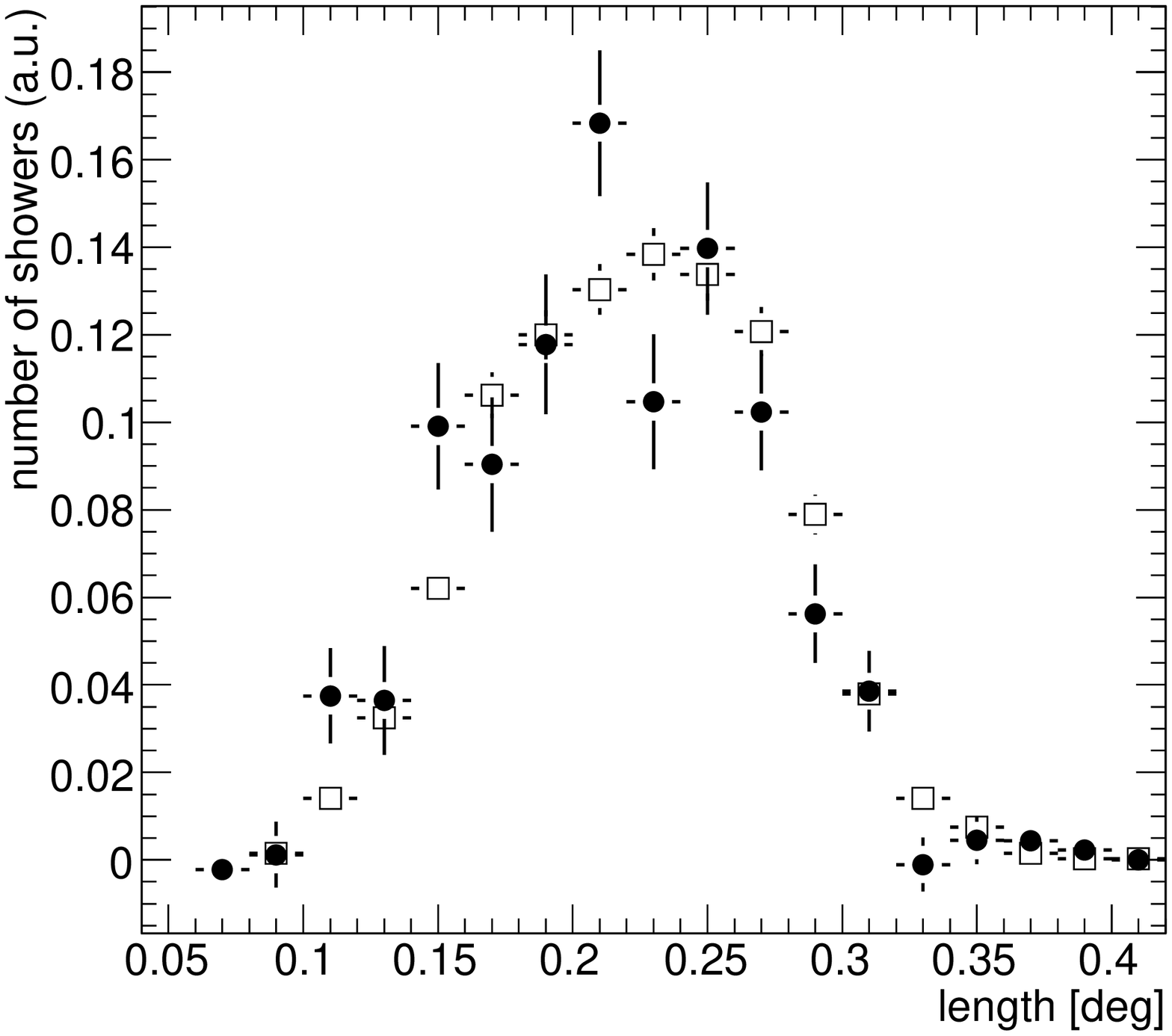}
\includegraphics*[width=0.41\textwidth]{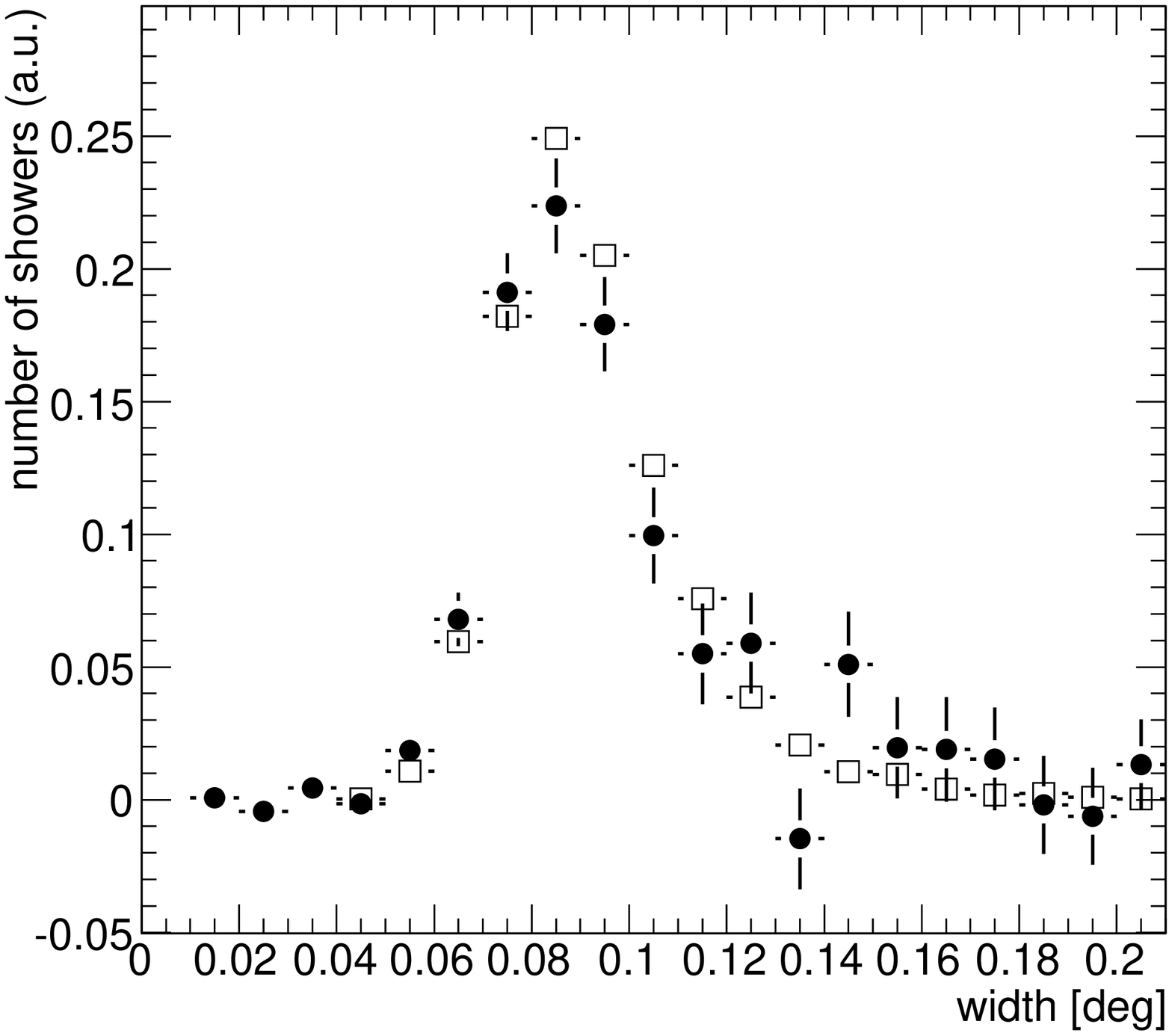}
\caption{\label {compare} Comparison of image parameter distributions from real
  data (closed symbols) and Monte Carlo calculations (open symbols). The
  parameters are described in the text.}
\end{center}
\end{figure}

Ideally, the best image parameter cuts can be derived from the Monte Carlo
simulations, which provide a data set of pure $\gamma$ rays with high
statistics. Simulations also provide the only method by which to estimate the 
primary energy of the incident $\gamma$ rays. For the simulations to be useful,
it is necessary to verify that the simulated images are an accurate
representation of the data. Figure~\ref{compare} shows the distributions of the image parameters $\alpha$,
$width$, $length$ and $distance$ for simulated $\gamma$ rays and for the
$\gamma$-ray excess observed in the real data. The agreement is good, indicating
that the simulations are valid.

\subsection{Crab Nebula Energy Spectrum}

\begin{figure}[h]
\begin{center}
\includegraphics*[height=8cm]{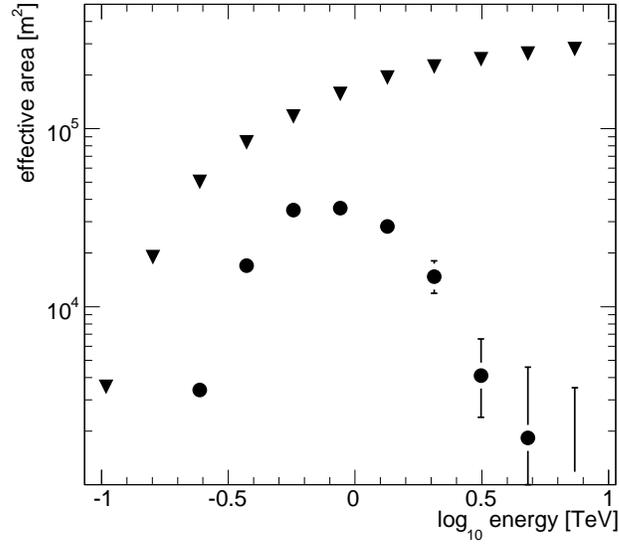}
\caption{ \label {effarea} Effective area of Telescope 1 after hard
  $\gamma$-ray selection cuts (filled circles) and $size$-dependent cuts (filled triangles).}
\end{center}
\end{figure}

The effective area of Telescope 1 after cuts is shown in Figure~\ref{effarea}.
The filled circles show the effective area for the hard $\gamma$-ray selection
cuts described above which reject most of the high energy events due to the
cut on $distance$. This is a problem which is well known from our experience
with the Whipple~10m telescope, and is solved by using $size$-dependent
$\gamma$-ray cuts \cite{Mohanty}, which do not provide
the same level of background rejection, but which take the dependence of the
image parameters on primary energy into account and retain $\sim80\%$ of the
$\gamma$-ray signal.

The maximum effective area after $size$-dependent cuts is $\sim2.8\times 10^5$
m$^{2}$. The energy threshold of a Cherenkov telescope is
conventionally defined as the position of the peak of the energy spectrum of
the source convolved with the effective area curve of the detector.
According to this definition, the threshold  is 150 GeV at trigger level, 160
GeV after applying $size$-dependent cuts, and 370 GeV after applying the hard
cuts described earlier.  

\begin{figure}[h]
\begin{center}
\includegraphics*[height=8cm]{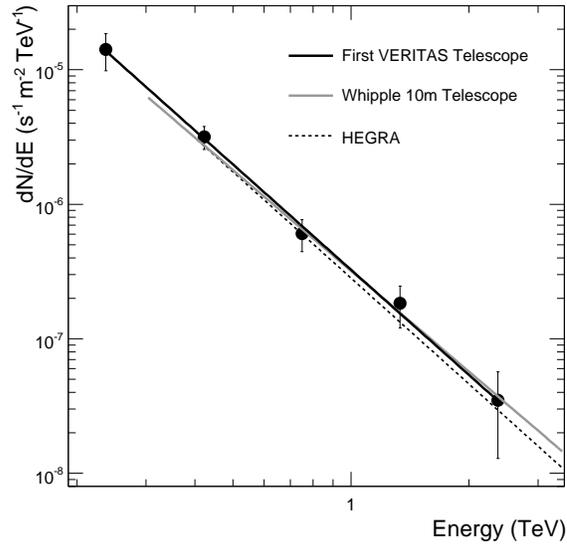}
\caption{ \label {crabspec} Energy spectrum of the Crab Nebula compared with
  earlier results from the Whipple $10\U{m}$ telescope \cite{Hillas98}
and HEGRA \cite{Aharonian04}.}
\end{center}
\end{figure}

In order to reconstruct the energy spectrum of $\gamma$ rays from the Crab
Nebula we have used the method outlined in Mohanty et al \cite{Mohanty}. $Size$-dependent cuts
are applied to the data, and the primary energy of the resulting $\gamma$-ray
excess events is estimated using a function which depends on the image $size$
and $distance$; the exact form of the function is derived from the Monte Carlo
simulations. Figure~\ref{crabspec} shows the resulting reconstructed energy spectrum of
the Crab Nebula.  A power-law fit to the data points gives a spectral index of $2.6\pm0.3$ and a
differential flux at $1\U{TeV}$ of $(3.26\pm0.9)\cdot10^{-7}$
m$^{-2}$s$^{-1}$TeV$^{-1}$ (statistical errors only).  This agrees well with
earlier measurements by other telescopes \cite{Hillas98,Aharonian04}.

\section{Discussion}

\begin{figure}[h]
\begin{center}
\includegraphics*[height=8cm]{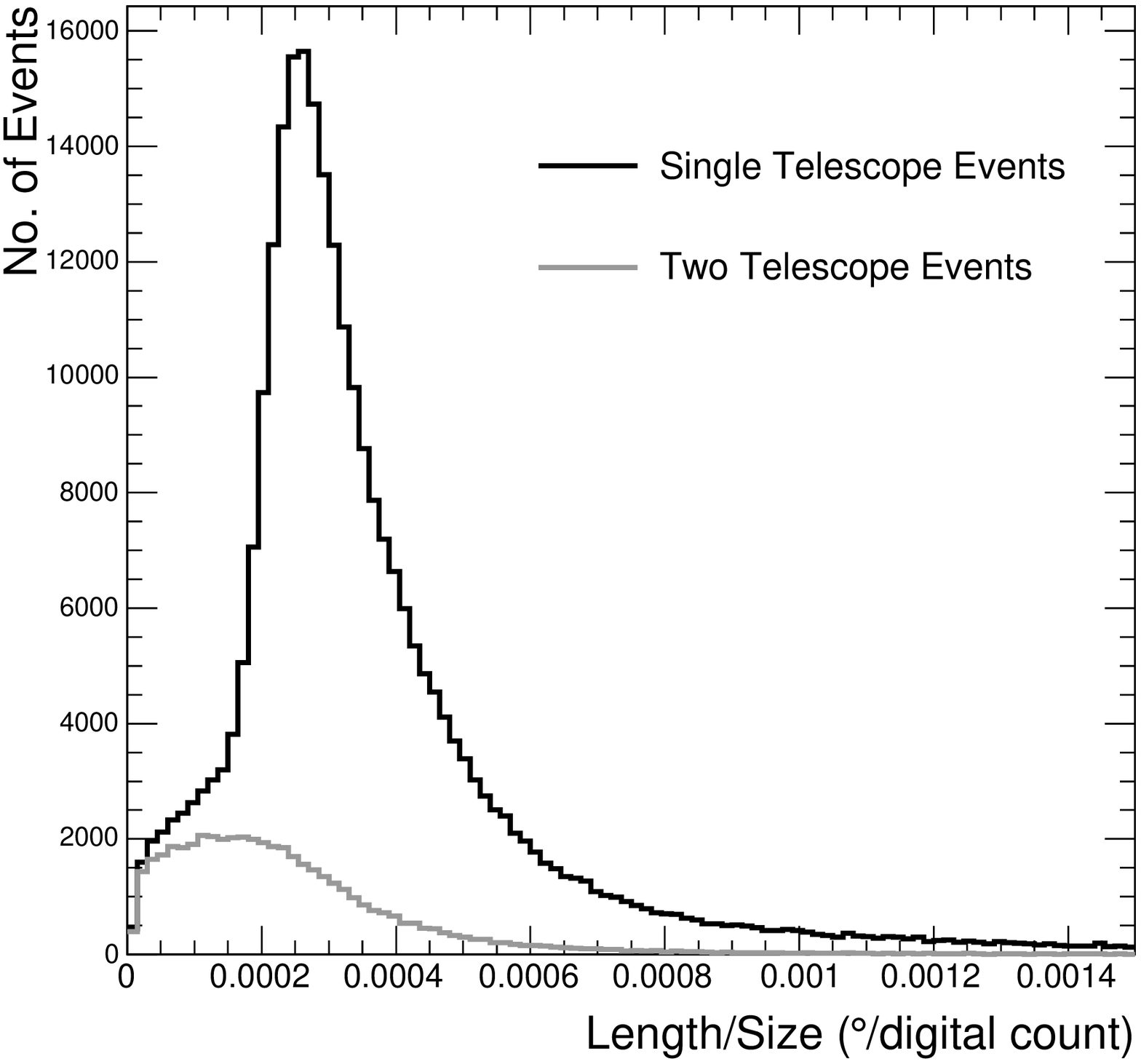}
\caption{ \label {los} $length$/$size$ distributions (parameters defined in
  the text). The single telescope
  histogram includes all events which trigger the telescope. The distinctive peak
  in this distribution is due mainly to local muon events. The two telescope
  distribution includes only events which triggered both telescopes within a
  $10\U{\mu s}$ time window. Events due to local muons are removed when this condition is imposed.}
\end{center}
\end{figure}

The first VERITAS telescope was operated throughout 2005, has met all
technical specifications and has detected a number of TeV $\gamma$-ray
sources. The sensitivity, at $\sim10\sigma$ for one hour of ON-source
observations of the Crab Nebula, is already appreciably better than the
Whipple $10\U{m}$ telescope (at $\sim7\sigma$ for one hour of Crab
observations); however, in order to provide the level of sensitivity required
from this generation of detectors, it is necessary to reject the muon
background at the hardware level. This can only be achieved by the
installation of further telescopes. The major mechanical components of all
four telescopes have now been delivered to the Mt. Hopkins base camp. The
second VERITAS telescope has recently been installed $85\U{m}$ away from the
first telescope on an East-West baseline. Figure~\ref{los} shows the
$length$/$size$ histogram for the first telescope for all events which
trigger, and for those which generate a trigger in both telescopes within
$10\U{\mu s}$ of each other. The distinctive peak due to local muon events in
the histogram of all events is clearly removed when the two-telescope trigger
is imposed. A true ``hardware'' two-telescope trigger is currently being
installed, which will correct for changing time delays due to the source
movement in the sky, and reduce the required coincidence time to
$\sim50\U{ns}$ or less. Only events which satisfy the two-telescope trigger
will be read out by the data acquisition system, reducing the data rate
dramatically and so enabling us to lower the individual telescope trigger
thresholds, and hence the energy threshold of the experiment. The angular
resolution for point source analysis with a single telescope is $0.17^{\circ}$
(Figure~\ref{crab2d}); additional telescopes will provide multiple views of each
shower, reducing the angular resolution by around a factor of two.

The agreement of the Monte Carlo simulations with observational data
illustrated here shows that we have a good understanding of the first VERITAS
telescope and that the design performance is being met. This being the case,
the predicted behaviour of the four-telescope array system is likely to be
reliable, indicating a flux sensitivity of
$3.4\times10^{-11}\UU{cm}{-2}\UU{s}{-1}$ at $100\U{GeV}$ for a 50 hour
exposure \cite{Fegan03}.
 
\section{Acknowledgments}
This research is supported by grants from the U.S.~Department of Energy, the
U.S. National Science Foundation,
the Smithsonian Institution, by NSERC in Canada, by Science Foundation Ireland and by PPARC in the UK.



\newpage


\begin{thebibliography}{00}




\bibitem{Galbraith53} Galbraith, W., \& Jelley, J.V., Nature, 171 (1953) 349
\bibitem{Weekes77} Weekes, T.C. \& Turver, K.E., Proc. 12th ESLAB Symp., Frascati, (1977) 279
\bibitem{Hillas85} Hillas, A.M. Proc. 19th ICRC, 3 (1985) 445
\bibitem{Weekes89} Weekes, T. C., et al. ApJ, 342 (1989) 379
\bibitem{Puhlhofer03} P\"{u}hlhofer, G. et al. Astropart. Phys., 20 (2003) 267
\bibitem{Lorenz04} Lorenz, E. et al. New Astron. Rev., 48 (2004) 339
\bibitem{Weekes02} Weekes, T.C. et al. Astropart. Phys., 17 (2002) 221
\bibitem{Hinton04} Hinton, J.A. et al. New Astron. Rev., 48 (2004) 331
\bibitem{Kubo04} Kubo, H. et al. New Astron. Rev., 48 (2004) 323
\bibitem{Aharonian05} Aharonian, F. et al. Science, 307 (2005) 1938
\bibitem{Wakely03} Wakely, S.P. et al. Proc. 28th ICRC, Tsukuba, Editors:
  T. Kajita, Y. Asaoka, A. Kawachi, Y. Matsubara and M. Sasaki, (2003) 2803
\bibitem{Gibbs03} Gibbs, K. et al. Proc. 28th ICRC, Tsukuba, Editors:
  T. Kajita, Y. Asaoka, A. Kawachi, Y. Matsubara and M. Sasaki, (2003) 2823
\bibitem{Davies57} Davies, J.M. \& Cotton, E.S. Journal of Solar Energy, 1
  (1957) 16
\bibitem{Hall03} Hall, J. et al. Proc. 28th ICRC, Tsukuba, Editors:
  T. Kajita, Y. Asaoka, A. Kawachi, Y. Matsubara and M. Sasaki, (2003) 2851
\bibitem{Bradbury02} Bradbury, S.M. \& Rose, H.J. Nucl. Instr. \& Meth. A, 481
  (2002) 521
\bibitem{Buckley03} Buckley, J.H. et al. Proc. 28th ICRC, Tsukuba, Editors:
  T. Kajita, Y. Asaoka, A. Kawachi, Y. Matsubara and M. Sasaki, (2003) 2827
\bibitem{Holder05} Holder, J. et al. Proc. 29th ICRC, Pune, (2005) in press.
\bibitem{Hanna02} Hanna, D. \& Mukherjee, R. NIM A, 482 (202) 271 
\bibitem{Vacanti94} Vacanti, G. et al. Astropart. Phys., 2 (1994) 1 
\bibitem{Humensky05} Humensky, T.B. et al. Proc. 29th ICRC, Pune, (2005) in
  press.
\bibitem{Roberts03} Roberts, M.D. et al. Auger GAP note, GAP-2003-010 (2003),
  http://www.auger.org
\bibitem{Shepherd05} Shepherd, N. et al. Proc. 29th ICRC, Pune, (2005) in
  press.
\bibitem{Lessard01} Lessard, R. et al. Astropart. Phys, 15 (2001) 1
\bibitem{Cogan05} Cogan, P. et al. Proc. 29th ICRC, Pune, (2005) in
  press.
\bibitem{Maier05} Maier, G. et al. Proc. 29th ICRC, Pune, (2005) in
  press.
\bibitem{Heck} Heck, D. et al., Report FZKA 6019, Forschungszentrum Karls\-ruhe (1998)
\bibitem{LeBohec} Duke, C. \& LeBohec, S. http://www.physics.utah.edu/gammaray/GrISU/
\bibitem{Hoerandel} H{\"o}randel, J. Astropart. Phys, 19 (2003) 193
\bibitem{Modtran} Kneizys et al. {\em The Modtran 2/3 report and lowtran 7 model}, Technical Report,
Ontar Corporation (1996)
\bibitem{Mohanty} Mohanty, G. et al., Astropart. Phys, 9 (1998) 15
\bibitem{Hillas98} Hillas, A.M. et al., Ap. J., 503 (1998) 744
\bibitem{Aharonian04} Aharonian, F. et al., Ap. J., 614 (2004) 897
\bibitem{Fegan03} Fegan, S. et al. Proc. 28th ICRC, Tsukuba, Editors:
  T. Kajita, Y. Asaoka, A. Kawachi, Y. Matsubara and M. Sasaki, (2003) 2847

\end{thebibliography}
\end{document}